\documentclass[onecolumn]{emulateapj}

\newcommand{\msun}{\mbox{$M_\odot$}}

\newcommand\be{\begin{eqnarray}}
\newcommand\ee{\end{eqnarray}}
\newcommand\lsim{\mathrel{\rlap{\lower3pt\hbox{\hskip1pt$\sim$}}
     \raise1pt\hbox{$<$}}} 
\newcommand\gsim{\mathrel{\rlap{\lower3pt\hbox{\hskip1pt$\sim$}}
     \raise1pt\hbox{$>$}}} 

\shorttitle{Mergers of Binary Compact Objects}
\shortauthors{Lee, Park, \& Brown}

\begin{document}

%
%

\title{Mergers of Binary Compact Objects}

\author{Chang-Hwan Lee, Hong-Jo Park}
\affil{Department of Physics, Pusan National University, Busan
609-735, Korea}
\email{clee@pusan.ac.kr}

\and

\author{Gerald E. Brown}
\affil{Department of Physics and Astronomy, State University of New
York, Stony Brook, NY 11794, USA}

%
%

%
%
%
%

\begin{abstract}
We work out the effects of hypercritical accretion, which transfers mass from the secondary to the primary (first-born) neutron star (NS) in a binary, showing that the mass of the primary would end up $\gsim 0.6\msun$ greater than the secondary NS in contradiction with the very nearly equal masses of the two NS's in binaries measured to date. We discuss that the primary NS evolves into a low-mass black-hole (LMBH) due to the hypercritical accretion.
Using a flat distribution in the mass ratio $q$ ($q\equiv M_{\rm secondary}/M_{\rm primary}$), favored by
\citet{Duq91}, we calculated a ratio of LMBH-NS  and
NS-NS systems to be $\sim 5$, in rough agreement with \citet{Pin06}. These authors emphasize the importance of ``twins",
which we discuss. The two NS's in the twins would be close in mass
and further increase the number of mergings.
\end{abstract}


%
%

\section{Introduction\label{intro}}

\citet{Bet98} suggested that mergers resulting in short-hard gamma-ray bursts would be
mainly those of LMBH-NS binaries, with those of NS-NS binaries
down by an order of magnitude from these. The lower number of the latter resulted
from the necessity that the two giant progenitors be within 4\% of each other in ZAMS
mass so that they burned He at the same time. Otherwise the first born pulsar would find
itself in the red giant envelope of the companion giant as it evolved and accrete enough
matter to go into a black hole (BH).
\citet{Bro95} had already estimated this to be true, based on \citet{Che93},
and proposed the special way that two giants burn He at the same time, in order to avoid the
red giant common envelope evolution of the first born pulsar.
If the two giants burn He at the same time, the two He envelopes are assumed to go into common envelope evolution, expelling the common envelope matter so that the helium envelope is lost from each star. \citet{Bra95} showed that there was not sufficient time for either of the stars to accrete the common envelope He, so that if the two He stars had to be nearly equal in mass, then their progenitor giants must also be. From the \cite{Sch92} models for the giant progenitors of the neutron stars we consider, the giants have to be within $\lsim 4\%$ of each other in mass in order to burn He at the same time.
Of the \cite{Bet98} merger rate of $10^{-4}$ yr$^{-1}$ in our Galaxy,
only $\sim 0.1$, or $10^{-5}$ yr$^{-1}$ were estimated to be those of binary NS's.
A recent detailed calculation by \citet{Dew96} gives a merger rate of
$0.1-12$ Myr$^{-1}$ for Brown's special scenario, the upper end of the calculation in
agreement with \cite{Bet98}.
The latter authors simply estimated that the remaining mergers would be of LMBH-NS
binaries since they calculated that when the pulsar went through
common envelope with the companion star it accreted $\sim 1\msun$ of matter, enough
to send it into a BH. The amount of accretion was corrected downwards $\sim 25\%$
by removal of an approximation of \citet{Bet98} by \citet{Bel02}.
In the present note we try to make a real calculation of the LMBH-NS binary, NS-NS
binary ratio.

\citet{Pin06} assembled evidence that ``Binaries like to be Twins".
They showed that a recently published sample of 21 detached eclipsing binaries
in the Small Magellanic Cloud can be evolved in terms of a  flat mass function containing
55\% of the systems and a ``twins" population with $q> 0.95$ containing the remainder.
All of the binaries had orbital period $P< 5$ days, with primary masses
$6.9 \msun <M_1 <27.3 \msun$.

The important role of twins is that the two giants are close enough in mass
\footnote{\citet{Pin06} used 5\% whereas we prefer 4\% as will be discussed.}
that in \citet{Bro95} scenario they can evolve into NS-NS binaries, whereas
if they are further apart in mass they will evolve into a LMBH-NS binary \citep{Che93,Bet98}.
Thus the twins may increase the number of NS-NS binaries. We suggest
that the resulting number of short hard gamma-ray bursts, which
result from the merging of the binaries, which to date are unable to
differentiate between the two species, may not be changed much, some
of the predicted large excess of LMBH-NS binaries appearing rather
as NS-NS binaries. However, because the latter are so much more easy
to observe, the role between what we see and what is present will be
tightened.

In Sec.~\ref{sec-twin} we shall show that evolution of binaries with
a flat mass evolution does, in agreement with \citet{Pin06}
 produce a ratio of (NS$+$BH)/(NS$+$NS) systems of 5, for the
population that does not contain twins. This is half the order of
magnitude  ratio found by \citet{Bet98}. The lower value results
from the fact that the secondary in the mass function is not
independent of the primary, as we shall see.  Taking the non twin
binaries to be 55\% of the total, the remaining 45\% twins,
\citet{Pin06} get a good fit to the detached binaries
measured in the Small Magellanic Cloud.

We point out that \citet{Bel02} in their simulation D2
in which the maximum NS mass is $1.5\msun$ and the mass proximity in
the progenitor binaries (to evolve NS's) is taken, like \citet{Pin06} to be 5\%, obtain a ratio of 4 for (BH$+$NS)/(NS$+$NS) and
would obtain the ratio of 5 had they used our 4\% proximity in
masses.

In short, there is general agreement amongst the authors quoted
above, except that it is not clear how many twins will be left once
selection effects are taken into account.
Historically large selection effects have been identified \citep{Gol03,Hog92}.
These will lower the number of twins found by \citet{Pin06}.

In Sec.~\ref{sec-bns} we summarize the calculations of maximum
NS mass obtained in a renormalization group calculation
expanding about the chiral restoration fixed point in the vector
manifestation of the Harada-Yamawaki hidden local symmetry theory
 \citep{Bro06}.
The result of $1.5\msun$ brings us into conflict with the recent
measurement of \citet{Nic05} of $2.1\pm0.2\msun$ for PSR
J0751$+$1807. The lower limit of this NS with 95\% confidence level
is $1.6\msun$. We evolve in our model of hypercritical accretion the
masses of pulsars and companions in NS-NS binaries, the result of
which is that the pulsars would have substantially greater masses
than their companions were a $2.1\msun$ NS star to be stable. We
find that our calculated distribution could be made consistent with
present observations, the most important of which require pulsar
mass to be no greater than $1.8\msun$. Thus we may have to raise our
calculated upper limit $0.2\msun$ but we believe that the lower end
of the mass measurement of J0751$+$1807 is favored.

\section{Evolution of Non-Twin Binaries}\label{sec-twin}

In the sample from the Small Magellanic Cloud, \citet{Pin06} found that the non-twin binaries and the twins are roughly equal (55 \% and 45 \%, respectively), and the non-twin population can be evolved
successfully with a flat mass distribution $dn/dm = {\rm constant}$.

Using a flat $q$ ($q\equiv M_2/M_1$)distribution, in which the IMF for the secondary (less massive) star is independent of its mass, $dN/dM_2={\rm Constant}/M_1$ for a given primary mass $M_1$, one can obtain the probability of having binaries with initial ZAMS masses within 4\%.
\begin{equation}
P_{\rm NS-NS} =
\frac{\int_{M_{\rm min}}^{M_{\rm max}} P_{\rm primary}(M_1)\; 
\left[ \int_{{\rm max}[M_{\rm min},0.96 M_1]}^{M_1} \left( dN/dM_2 \right) dM_2 \right] dM_1}
{\int_{M_{\rm min}}^{M_{\rm max}} P_{\rm primary}(M_1)\; 
\left[ \int_{M_{\rm min}}^{M_1} \left( dN/dM_2 \right) dM_2 \right] dM_1}
\end{equation}
where $M_{\rm min}$ and $M_{\rm max}$ are the lower and upper limits in ZAMS masses for which neutron star can be formed after supernova explosion, and $P_{\rm primary}$ is the mass function for the primary in binaries.
We use the Salpeter mass function as the IMF for the primary,
\begin{equation}
P_{\rm primary} (M_1) \propto M_1^{-2.35} \label{eq1}.
\end{equation}
Note that the lower limit of $0.96 M_1$ in the numerator comes from the fact that the mass of the secondary star has to be within 4\% in ZAMS mass in order to make the double pulsar.

\begin{figure}[t]
\epsscale{.7}
\plotone{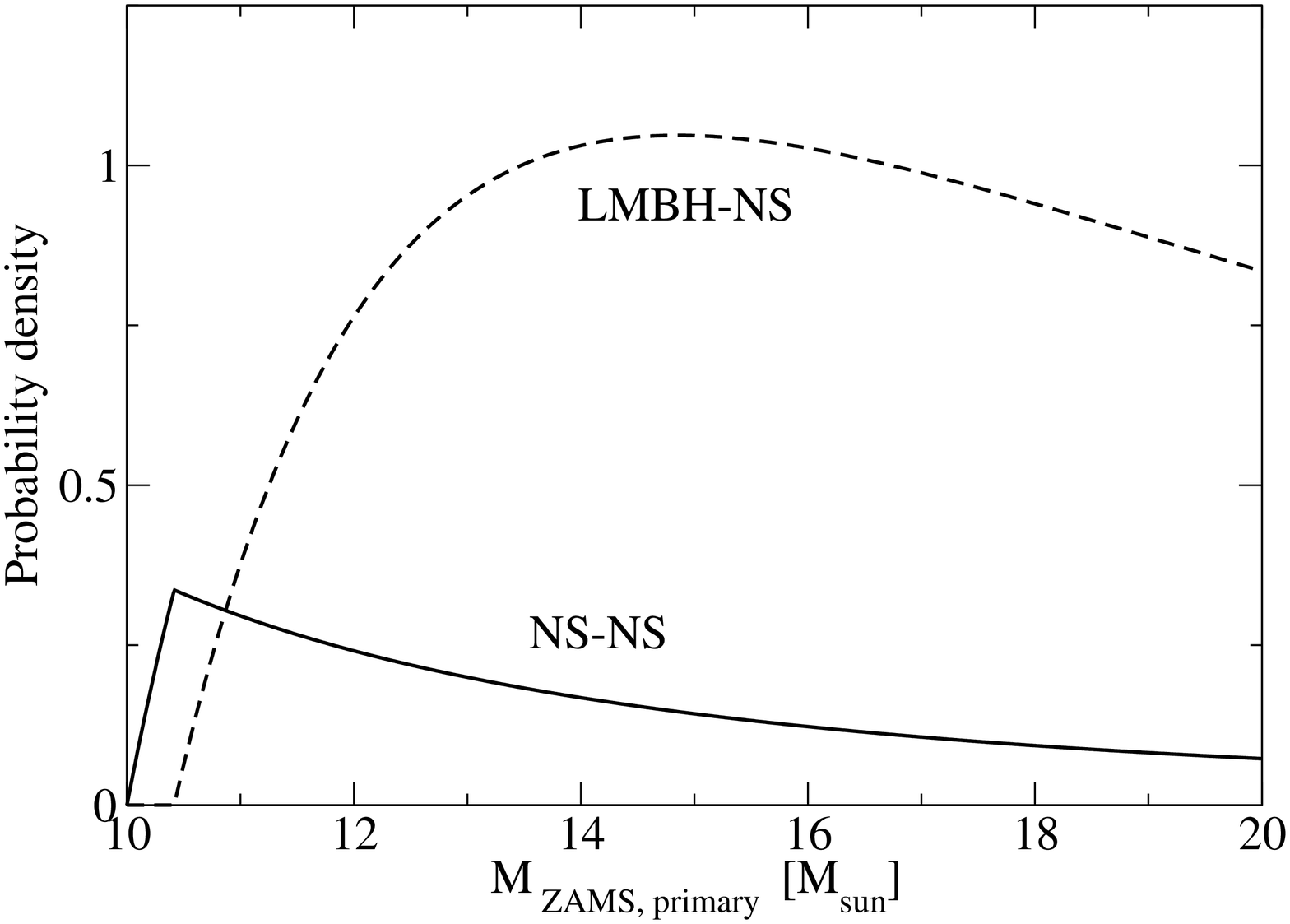}
\caption{Probability density of the formation of NS-NS and NS-LMBH binaries. The x-axis is the ZAMS mass of the progenitors of primary stars in compact binaries.
Here we assumed that the accreted pulsars, whose masses are bigger than $1.8\msun$, went into black holes as in Fig.~\ref{fig2}. }\label{fig1}
\end{figure}

There is no clear understanding on the relation between the pulsar masses and their progenitors.
We gave arguments that the
Hulse-Taylor pulsar 1913$+$16 was evolved from a $20\msun$ ZAMS
giant \citep{Bet07}. (See \citealt{Bur86}.) The
companion in J1756$-$2251 of $1.18^{+0.03}_{-0.02}$ is the lowest
measured mass in NS binaries, and we take it to have originated from
a $10\msun$ giant. Our interval in which binary NS's can be formed
is taken to be $10-20\msun$ as in \citet{Bet98}.

From eq.~\ref{eq1}, with given assumptions, we finds that NS-NS binaries should be formed 16\% of the time, resulting in 5.3 times more NS-LMBH binaries, which are harder to be observed, than NS-NS binaries.
The estimated probability distribution is summarized in Fig.~\ref{fig1}.
From the figure, one can see that more NS-NS binaries can be formed with lower ZAMS mass primary stars, while LMBH-NS binaries mainly come from binaries with higher ZAMS mass primary stars (mainly because the possibility of forming NS stars out of secondary stars are higher).

By requiring stars to be within 5\%
of each other in mass to burn He at the same time,
\citet{Bel02} get a ratio of 4
to 1 for LMBH-NS binaries over NS-NS binaries. From eq.~(\ref{eq1}) with 5\%, we obtained 20\% chance of making NS-NS binaries, giving 4 times more
LMBH-NS binaries than NS-NS binaries, which are consistent with \citet{Bel02} within 10\% uncertainty.
So not only does the flat distribution fit the
non-twin binaries in the Small Magellanic Cloud but it reproduces
Belczy\'nski's result for the same assumed proximity of 5\% in mass.
(Note that there can be extra uncertainties in this direct comparison because different evolutionary phases
in different approaches can alter the final ratio.)

The hypercritical accretion plays a central role in the result of \citet{Bel02} as well as ours and the agreement between our results and their is perhaps better than would be expected in such a complex situation. We know hypercritical accretion to be present, because the trapping of photons and subsequent adiabatic inflow is based on the same physics, but with quite different parameters, as the trapping of neutrinos in the infall of matter in the collapse of large stars.
A rough estimate of the correction to the original \citet{Bet95} result, which neglected the compact object mass compared with the varying giant mass, gives the $\sim 25\%$ decrease to get from the \citet{Bet95} results to those of \citet{Bel02}.

The supernova explosion in a binary of two He stars to produce a double neutron star were carried out by \citet{Bet98} in a simple schematic way with a simple physical description of the survival rates, i.e., of the binary staying together after each explosion. Computer calculations by \citet{Wet96} showed the \citet{Bet98} assumption to be accurate to $\le 10\%$, slightly greater than with the formula of \citet{Por98}. In principle the merging of neutron stars and black holes could be very different because initially the mass transfer from neutron star into black hole is stable, so the black hole can ``bounce" away from the neutron star \citep{Por98}. The regularities in this are usually assumed to be broken up by the gravitational interactions, etc. and the binaries are assumed to merge by some kind of Roche Lobe overflow. Thus, although the manner of merging may be quite different for binary neutron stars and a neutron star and low-mass black hole, we believe that the ratio of mergings will be given by the population synthesis. 

\section{Binary Neutron Stars vs LMBH-NS Binaries} \label{sec-bns}

Kaon condensation has been suggested as a possible mechanism which can reduce the
maximum mass of neutron stars \citep{Bet95}.
Recently, a theoretical calculation of strangeness
condensation has been carried out by expanding about the fixed point
of the Harada-Yamawaki Vector Manifestation \citep{Bro06}. In
the strangeness condensation, which takes place at a density not
much less than that of chiral restoration, the fixed point, the
in-medium $K^-$ mass $m_{K^-}^\star$ has been brought down to the
value of the electron chemical potential $\mu_e$, so that the highly
degenerate fermionic electrons collapse into a zero-momentum Bose
condensate of $K^-$ mesons. Only mesons and constituent quarks are
left at densities close to the chiral restoration density $n_{\chi
SR}$. Expanding about the fixed point eliminates the uncertainties
introduced by many investigators about the role of strange mesons
and the role of the explicit symmetry breaking in the strangeness
sector.
%
%
%
%
In fact, \citet{Bro06} show that the $K^-$ meson behaves like
the nonstrange mesons, $\rho, \pi, \sigma, A_1$ in that its mass
goes to zero as $n\rightarrow n_{\chi SR}$. However, the criterion
for kaon condensation is much more easily fulfilled, because the
{\it in-medium} mass $m_{K^-}^\star$ has only to come down a little
more than half way to zero; i.e., to the electron chemical potential
$\mu_e$, for the phase transition to take place.

In \citet{Bro06} interactions between the bosons in the
condensate were ignored. They would not be expected to be large
because the $K^-$ mesons at condensation have a density of only
$\sim 0.05$ fm~$^{-3}$, an order of magnitude less than the neutron
density at kaon condensation. Thus, the $K^-$ in the condensation
are very diffuse. We can make this more quantitative by calculating
their interaction. The optical model potential from the other
$K^-$-mesons by a given $K^-$ is
\be
V_{\rm opt} =\frac{4\pi a}{(M_{K^-})_{\rm reduced}} n_{K^-}
\ee
where $a$ is the $K^-$-$K^-$ scattering length and the reduced mass
is used. The $a$ can be obtained\footnote{We are grateful to
Christof Hanhart for this information.} from the $I=2$ $\pi$-$\pi$
scattering length by changing $M_\pi$ to $M_{K^-}$, giving
\be
a=-\frac{M_{K^-}}{16\pi f_\pi^2}.
\ee
Thus
\be
V_{\rm opt} = n_{K^-}/2 f_\pi^2
\ee
where $f_\pi$ is the pion decay constant $f_\pi \sim 90$ MeV or one
can put in higher-order effects and use $f_K\simeq 108$ MeV. The
latter, together with
\be
n_{K^-} = 0.048 {\rm fm}^{-3}
\ee
at kaon condensation would give
\be
V_{\rm opt} = 16 {\rm MeV},
\ee
sufficiently negligible compared with the $\sim 250$ MeV in going
from $M_{K^-}$ to $M_{K^-}^\star \sim M_{K^-}/2$ at kaon
condensation. Thus, we stick by our calculation in Brown et al.
(2006).
None the less, in the light of the measurement of the mass of
PSR~0751$+$1802 \citep{Nic05}, it is very interesting to see
how far the maximum neutron star mass can be pushed up without the
pattern of masses in neutron star binaries running into conflict
with observation.
Later we shall see that  there is strong evidence of a NS more
massive than $1.5\msun$ existing, although we discuss that the limit
can be extended to $1.8\msun$ without destroying the pattern of
accurately measured NS's to date.

In the original \citet{Bet98} calculation of hypercritical
accretion the common envelope evolution was carried out
analytically, using Newton's and Kepler's laws and the coefficient
of dynamical friction $c_d=6$. The equations could be solved
analytically only if the NS mass $M_{NS}$, the smallest of the
masses, was set equal to zero. This approximation was removed by
\citet{Bel02}. Incorporating their corrections we find
that an originally $1.4\msun$ pulsar will have accreted $0.75\msun$
in the H red giant stage. This is enough to send it into a black
hole, and this is our correction of the conventional model which
ignores the accretion.

\begin{figure}[t]
\epsscale{.7}
\plotone{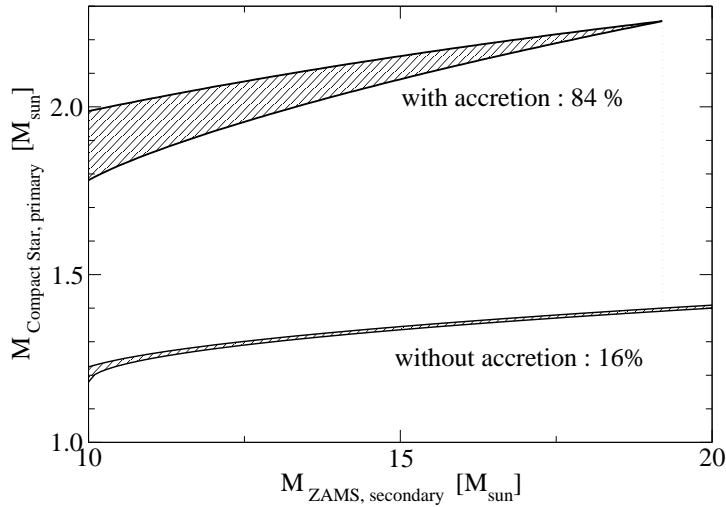}
\caption{Masses of primary compact stars with and without accretion during H red giant stage of secondary star (refer Table~\ref{tab2} for the details). Note that the 84\% corresponds to $M_{\rm Compact\; Star, primary}> 1.8\msun$. There is uncertainty in the final primary compact star masses due to the extra mass accretion, $\sim 0.1 \msun - 0.2\msun$, during He giant stage. This may increase the primary compact star masses  for both `with-' and `without-accretion'. Note that with our maximum neutron star mass of $1.8\msun$, all primary compact stars with accretion would go into low-mass black holes.
}\label{fig2}
\end{figure}

Lower mass He stars have to burn hotter than the higher mass ones
because of the loss of energy through the surface. Ordinarily the
range of $M_{He}=2-4\msun$ is chosen as a limit below which they
begin a red giant phase. We believe from the equality of masses in
B1534$+$12 that not much He can have been transferred in the pulsar,
He-star stage of the evolution. We, therefore, introduce He red
giant evolution below the ZAMS mass of $\sim 15\msun$ for this
binary.
The double pulsar with companion mass
$1.25\msun$ certainly has been recycled by mass deposited by its
companion during its He red giant stage. Whereas we calculate the
hypercritical accretion in He red giant to be $0.2\msun$ \citep{Bet07} the actual difference in J0737$-$3039 between pulsar and
companion mass is $0.09\msun$. However, the $0.2\msun$ is roughly
correct for J1756$-$2251. The He wind is quite uncertain. In fact,
\citet{Bel01} derived a way in which no mass was
accreted, similar to the \citet{Bro95} double He star scenario, which
expelled the H envelopes, except that now the He envelopes from the
two stars meet and are expelled.
For other possible evolutionary scenarios of NS-NS binaries, please refer
\citet{Yam93,Fry97,Arz99,Fra02,Dew03,Wil04,Tho05}.

\begin{table}
\begin{center}
\caption{Calculated accretion onto the pulsar during H red giant stage of secondary star. Masses are given in $\msun$. Note that for a given secondary star mass $M_{\rm ZAMS,secondary}$, the primary star can have masses in the range $M_{\rm ZAMS, secondary} < M_{\rm ZAMS, primary} < M_{\rm max}=20\msun$. However, in order to have mass accretion during H red giant stage, the mass difference has to be
bigger than 4\%. So, in the second column, for the estimation of the lower limit of the accretion, we take $M_{\rm ZAMS,primary}^{\rm min} = M_{\rm ZAMS, secondary}/0.96$. In the third column, we take $M_{\rm ZAMS,primary}^{\rm max} = 20 \msun$ for the estimation of the upper limit.
$M_{\rm C,i}$ is the initial primary compact star mass and $M_{\rm C,f}$ is the
final mass following accretion.  The He core mass of giant star
is assumed to be $M_{\rm He} = 0.08 (M_{\rm Giant}/\msun)^{1.45}
\msun$.
Note that there can be extra mass accretion during He giant stage, $\Delta M_{\rm C,f} = 0.1 \msun - 0.2\msun$, therefore, $M_{\rm C,f}$ in this table can be treated as a lower limit.
} \label{tab2}

\vskip 4mm

\begin{tabular}{|cc|ccc|ccc|}
\hline
\multicolumn{2}{|c|}{Secondary star}
   &\multicolumn{3}{c|}{Primary star; Minimum mass}
                   &\multicolumn{3}{|c|}{Primary star; Maximum mass} \\
$M_{\rm ZAMS}$ & $M_{\rm NS}$ &
$M_{\rm ZAMS}^{\rm min}$ & $M_{\rm C,i}$ & $M_{\rm C,f}$ &
$M_{\rm ZAMS}^{\rm max}$ & $M_{\rm C,i}$ & $M_{\rm C,f}$ \\
\hline
20 & 1.40                 &  --- & ---  & --- & --- & --- & --- \\
19 & 1.39 (B1913$+$16)    & 19.8 & 1.40 & 2.25 & 20 & 1.40 & 2.25 \\
18 & 1.38                 & 18.8 & 1.39 & 2.21 & 20 & 1.40 & 2.23 \\
17 & 1.36                 & 17.7 & 1.37 & 2.17 & 20 & 1.40 & 2.20 \\
16 & 1.35                 & 16.7 & 1.36 & 2.13 & 20 & 1.40 & 2.18 \\
15 & 1.34 (B1534$+$12)    & 15.6 & 1.35 & 2.08 & 20 & 1.40 & 2.15 \\
14 & 1.32                 & 14.6 & 1.33 & 2.03 & 20 & 1.40 & 2.12 \\
13 & 1.30                 & 13.5 & 1.31 & 1.98 & 20 & 1.40 & 2.09 \\
12 & 1.28                 & 12.5 & 1.29 & 1.93 & 20 & 1.40 & 2.06 \\
11 & 1.25 (J0737$-$3039B) & 11.5 & 1.26 & 1.86 & 20 & 1.40 & 2.02 \\
10 & 1.18 (J1756$-$2251)  & 10.4 & 1.22 & 1.78 & 20 & 1.40 & 1.99 \\
\hline
\end{tabular}
\end{center}
\end{table}

Using the four accurately measured companion masses in relativistic binary NS's
\footnote{We could easily also fit in 2127$+$11C, but it is generally thought to have been
made by recombination of NS's formed in different binaries.} we construct the final pulsar masses as in Fig.~\ref{fig2} and Table~\ref{tab2}.  Our strategy is to calculate the mass accreted on the primary pulsar in the H red giant
stage of secondary star. There can be an extra accretion in the He red giant stage, which will increase the final primary compact star mass.  We also neglected mass
transfer by He winds in the pulsar, He star binary which precedes the NS-NS stage,
but this is likely to be small because of the propeller effect \citep{Fra02}.
For the relation between the pulsar and progenitor masses, we assumed
\begin{equation}
M_{\rm Pulsar} = (1.40 \msun -1.18\msun)\times\left[\frac{M_{\rm ZAMS}-10\msun}{20\msun-10\msun}\right]^{0.5} +1.18 \msun.
\label{eq-pul}
\end{equation}
Note that this approximation is based on the known pulsar masses, as in Table~\ref{tab3a}, and the pulsar and progenitor mass relations, which we discussed above and marked in Table~\ref{tab2}.

Using the \citet{Bet98} hypercritical accretion corrected by \citet{Bel02}
we arrive at Fig.~\ref{fig2} and Table~\ref{tab2}, giving the estimated final primary compact star masses.
Note that there can be extra mass accretion during He giant stage, $0.1 \msun - 0.2\msun$, therefore, final compact star masses in Fig.~\ref{fig2} and Table~\ref{tab2} can be treated as a lower limit.

\begin{table}
\caption{Compilation of NS-NS binaries \citep{Lat07}. Individual references are
(a) \citealt{Tho99}, (b) \citealt{Lyn04}, and (c) \citealt{Fau04}.}
\label{tab3a}
\begin{center}
\begin{tabular}{lllc}
\hline
Object      & Mass ($\msun$) & Companion Mass ($\msun$) & References \\
\hline
J1518$+$49 & 1.56$^{+0.13}_{-0.44}$ & 1.05$^{+0.45}_{-0.11}$ & (a) \\
B1534$+$12 & 1.3332$^{+0.0010}_{-0.0010}$ & 1.3452$^{+0.0010}_{-0.0010}$ & (a) \\
B1913$+$16 & 1.4408$^{+0.0003}_{-0.0003}$ & 1.3873$^{+0.0003}_{-0.0003}$ & (a) \\
B2127$+$11C & 1.349$^{+0.040}_{-0.040}$ & 1.363$^{+0.040}_{-0.040}$ & (a) \\
J0737$-$3039A  & 1.337$^{+0.005}_{-0.005}$
       & 1.250$^{+0.005}_{-0.005}$ (J0737$-$3039B)  & (b) \\
J1756$-$2251  & 1.40$^{+0.02}_{-0.03}$ & 1.18$^{+0.03}_{-0.02}$ & (c) \\
\hline
\end{tabular}
\end{center}
\end{table}

In Table~\ref{tab3a} we list the known relativistic NS-NS binaries.
The masses of the two least massive binaries J0737$-$3039A, B and J1756$-$2251
are slightly different because of mass transfer during the He red giant stage,
but certainly the near equality of masses within a binary is completely
different from that in Table~\ref{tab2}. Thus, we conclude that the pulsar
must have evolved into a BH in most of the cases, leaving only those from
double He star burning; i.e., those in which the two NS's are within 4\% of
each other in mass.

We note that the masses of NS's evolved in binaries with white
dwarfs by \citet{Tau99} are similar to those in the
column labeled $M_f$ in our Table~\ref{tab2}, 2/3 of their final
masses being greater than $2\msun$. Indeed, their NS mass for
J0751$+$1807 is $2.16 \msun$, to compare with the measured mass of
$2.1\pm 0.2\msun$ \citep{Nic05} and companion mass is $0.181
\msun$ to compare with the \citet{Nic05} $0.191 \pm 0.015 \msun$. The
\citet{Nic05} values here are with 68\% confidence, whereas they are
$2.1^{+0.4}_{-0.5} \msun$  for the NS and $0.191^{+0.033}_{-0.029}$
for the white dwarf with 95\% confidence.

\citet{Bet95} obtained a maximum mass for the compact core of
1987A of $1.56\msun$, assuming that it had gone into a black hole,
from the $\sim 0.075 \msun$ of Ni production.
From our result in Fig.~\ref{fig2}, the upper limit on the maximum NS mass
can be pushed up to $\sim 1.8\msun$.
Our maximum NS mass, $\lsim 1.8\msun$, cannot
be reconciled with the observations. However, it is noteworthy that
only one of the dozen or so NS, white-dwarf binaries observed
exceeds our limit, whereas most of \citet{Tau99} do.
The \citet{Nic05} error will be improved with further observation. It
will be interesting to see how their results evolve.

%
%

\section{Common Envelope Evolution in NS-COWD
Binaries} \label{sec-cowd}

Our scenario that in NS-NS binary evolution the pulsar goes into a
BH if common envelope evolution takes place while the companion is
red giant has been challenged in the literature, because it is known
that a number of NS, carbon-oxygen  white dwarfs (COWD) PSR
1157$-$5112, J1757$-$5322, J1435$-$6100, J1454$-$5846, J1022$+$100,
and J2145$-$750 have survived common envelope evolution, from the
fact that the NS is strongly recycled. At least two unrecycled
pulsars, B2303$+$46 and J1141$-$6545 have been observed.

The behavior of the pulsar magnetic field is crucial for the observability times of NS's. \citet{Van94b} has pointed out that NS's formed in strong magnetic fields $10^{12} - 5\times 10^{12}$ Gauss, spin down in a time $\tau_{\rm sd}\sim 5\times 10^6$ years and then disappear into the graveyard of NS's. The pulsation mechanism requires a minimum voltage from the polar cap, which can be obtained from $B_{12}/P^2 \gsim 0.2$ with $B_{12}=B/10^{12}~{\rm G}$ and $P$ in seconds.
As a result, the relative time that the pulsar is observable is proportional to the observability premium $\Pi$
\citep{Wet96}, 
the ratio of observability for a pulsar with magnetic field $B$ to one of $B=10^{12}$ G,
\be
\Pi = 10^{12} {\rm G} /B
\label{premium}
\ee
where $B$ is the magnetic field strength of the pulsar, and the spin down depends only on $B$.
%
%

Since the magnetic fields $B$ of recycled pulsars have been brought down substantially by accretion, presumably from their companions, as found phenomenologically by \citet{Taa86}, the pulsar that have had a lot of recycling, and, therefore lower $B$, can be observed for a much longer time than ``fresh" pulsars that have not been recycled.

\begin{table}
\begin{center}
\caption{Observability premiums of unrecycled and recycled pulsars.
}
\label{tab3}
\vskip 2mm
\begin{tabular}{cc}
\hline
Unrecycled  Pulsar & $\Pi$ \\
\hline
            B2303$+$46 & 1.26 \\
            J1141$-$6545 & 0.77 \\
\hline
Recycled Pulsar & \\
\hline
          J2145$-$0750 & 1667 \\
          J1022$+$1001 & 1190 \\
          J1157$-$5112 & 159 \\
          J1453$-$5846 & 164 \\
          J1435$-$6100 & 2127 \\
          J1756$-$5322 & ? \\
\hline
\end{tabular}
\end{center}
\end{table}

We plot the observability premium of the unrecycled and recycled pulsars separately in
Table~\ref{tab3}.
We see from Table~\ref{tab3} that the chance of observing a recycled binary, because of
the longer lifetime of the pulsar, is at least two orders of magnitude greater than
observing an unrecycled pulsar. We interpret this as meaning that almost all of the
recycled pulsars go into LMBH-NS binaries in common envelope evolution, but there
are a few exceptions.

The discussion by \citet{Van94a} of 2145$-$0750 is illuminating as to how
the NS binary might survive. He requires an efficiency for expelling the envelope of a
factor of several greater than usual in common envelope binary evolution \citep{Web84}, interpreting this
as meaning that agencies other than the dynamical friction are helping to expel the envelope.
The use of efficiencies greater than unity is often employed in order to include phenomenologically additional effects that help the binary survive common envelope evolution \citep{Ibe93,DeK90,Yun93}.
We know that in the asymptotic giant branch the envelope is lost, in any event, as the
originally main sequence star evolves. Possibly the strong gravitational field of the
NS helps in expelling the envelope. In any case it is clear that the most likely fate of a
NS-COWD which has gone through common envelope evolution is that the NS accretes enough
matter to go in a LMBH.

Our assumption is that survival probabilities for the double neutron star formation and NS-LMBH formation are the same. Unfortunately, this cannot be checked because it has not been possible to distinguish in the short-hard gamma-ray bursts  between the two types of events \citep{Nak07}.

\section{Conclusion} \label{sec-con}

The work of \cite{Pin06} on the 21 detached binaries in
the Small Magellanic Cloud suggests that there are about as many
giant stars in twins as in the non-twin population, although the
number in twins will be decreased by selection effects which were
not taken into account. The twins are within 5\% of each other in
ZAMS mass and will therefore burn He at the same time so that
binaries with sufficiently massive stars ($\gsim {\rm ZAMS} \ 10
\msun$) will evolve into NS-NS binaries.

These will add, in the merging of compact binaries to the \citet{Bet98} LMBH-NS
binary mergings. Neglecting the contribution from twins, these are shown to increase the
NS-NS binary mergers by a factor of 6, down by a factor of 2 from the rough estimate of \cite{Bet98}, when evolved with a flat mass function for the companion
star. The latter is shown to favor binaries of low mass NS, like the NS-NS binary,
because the companion mass is correlated with the pulsar mass by its necessity of
being less massive.

We reconsidered the maximum mass determination of $1.5\msun$ for a NS, first proposed
by \citet{Bet95} and \citet{Bro94}, in the light of the measurement
of the $2.1\pm 0.2\msun$ mass in PSR 0751$+$1807.
We show that our maximum NS mass could be raised to $1.8\msun$, within 95\%
confidence value of \citet{Nic05}, but that increasing it more would lead
to the pulsars of higher mass than $1.8\msun$ remaining stable following hypercritical
accretion during the red giant evolution of the companion (which is necessarily less
massive than the pulsar at formation times of the two NS's). Thus the pattern of NS-NS
binaries would chiefly consist of pulsar which are $\sim 50\%$ massive than companion NS, which is
not seen.



\section*{Acknowledgments}

C.H.L. was supported by the Korea Research Foundation Grant funded
by the Korean Government(MOEHRD, Basic Research Promotion Fund)
(KRF-2005-070-C00034). G.E.B. was supported in part by the US
Department of Energy under Grant No. DE-FG02-88ER40388.


\end{document}